\documentclass[5p,twocolumn,number,times]{elsarticle}

\newcommand{\bear}{\begin{eqnarray}}
\newcommand{\eear}{\end{eqnarray}}
\newcommand{\be}{\begin{equation}}
\newcommand{\ee}{\end{equation}}
\newcommand{\beqn}{\begin{eqnarray}}
\newcommand{\eeqn}{\end{eqnarray}}
\newcommand{\beqnn}{\begin{eqnarray*}}
\newcommand{\eeqnn}{\end{eqnarray*}}

\def\vf{\varphi}

\journal{Physics Letters A}
 
\begin{document}

\begin{frontmatter}
\title{Tunneling of slow quantum packets through the high Coulomb barrier}

\author{A.V.~Dodonov}
\ead{adodonov@fis.unb.br}

\author{V.V.~Dodonov\corref{cor1}}
\ead{vdodonov@fis.unb.br}

\cortext[cor1]{Corresponding author. Tel.: (55)(61) 31076097; Fax: (55)(61)33072363}
\address{
Instituto de F\'{\i}sica, Universidade de Bras\'{\i}lia,
Caixa Postal 04455,
70910-900 Bras\'{\i}lia, DF, Brazil
}

\begin{abstract}
We study the tunneling of slow quantum packets through a high Coulomb barrier.
We show that the transmission coefficient can be quite different from the standard expression
obtained in the plane wave (WKB) approximation (and larger by many orders of magnitude), 
even if the momentum dispersion is much smaller than the mean value of the momentum.

\end{abstract}


\end{frontmatter}

\section{Introduction}

One of the most striking properties of the quantum world is  the {\em tunneling effect\/}
\cite{Gamow,Gurney29}.
The well known quasiclassical formula for the transmission probability through the potential barrier $U(x)$ reads
\be
D \sim \exp\left(-2\int_1^2\sqrt{2m\left[U(x) -E\right]}dx/\hbar\right),
\label{Gamow}
\ee
 where $\hbar$ is the Planck constant, $m$ is the mass of particle and $E$ is its energy.
But formula (\ref{Gamow}) was derived for idealized quantum states with a well defined energy,
represented by the plane wave having infinite extension in the coordinate space. A more realistic
approximation is to consider the tunneling of wave packets. 
Then one has to average the coefficient $D(p)$ over the wave function $\vf(p)$ of the quantum packet
in the momentum space, taking into account only the
plane waves going in the direction of the barrier:
\be
T= \int_0^{\infty} D(p)|\vf(p)|^2 dp.
\label{Dav}
\ee
This was done  for very narrow superpositions in the momentum space as far back as in Ref.
\cite{MacColl32}. 
\footnote{ Formula (\ref{Dav}) seems quite obvious. For its justification see, e.g., \cite{AD02}.}
 On the other hand, it was shown later that various new effects can arise
in scattering and tunneling of packets which are not extremely narrow in the momentum space
(in particular, narrow in the coordinate space)  \cite{DKM-PLA96,Kadom,Ignat,AD04}.
Here we consider this problem for the Coulomb potential barrier, which has numerous applications,
especially for the fusion and radioactive decay phenomena. 
We make emphasis on the transmission probabilities of {\em slow particles}, because this
regime attracted significant attention in attempts to explain experimental data related
to low energy nuclear reactions \cite{Balan98,Storms10,Vysot10,Vysot12a,Vysot13,Ivlev13}.
It seems obvious that the spread of the packet in momentum space should result in the increase
of the barrier transparency, due to the enhanced contribution of the plane wave components with
high values of momenta. What is not so obvious (at least, unexpected), it is the fact that 
even small dispersions of the momentum can result in increase of the transmission coefficient
by many orders of magnitude. This is the motivation for writing this article.

\section{Tunneling of wave packets}

We confine ourselves to the idealized
barrier in a single space dimension, $U(x)=Ze^2/x$ for $x>0$ (and $U(x)=0$ for $x<0$). 
The integral (\ref{Gamow}) is well known
in this case (we omit a possible pre-exponential factor): 
\be
D(p)= \exp\left( -a/p\right), \qquad 
a=2\pi Ze^2 m/\hbar,
\label{Dp-id}
\ee
where $p$ is the linear momentum. We know that the standard WKB approximate formula (\ref{Gamow})
needs modifications for the long-range potentials, such as the Coulomb one, but this is not
essential for our purposes. Let us simply assume that $D(p)$ is given by Eq. (\ref{Dp-id}), and
let us see what can happen for different wave packets.

If the packet is concentrated near the mean value $p_0$ and  its spread can be well
characterized by the momentum variance $\sigma_p$, then integral (\ref{Dav}) gives the
value close to $D(p_0)=\exp(-A)$ under the condition
\be
A\sqrt{B} \ll 1,
\label{cond}
\ee
where
\be
A= \frac{a}{p_0}=\frac{2\pi Ze^2}{\hbar v_0}, \qquad
B= \frac{\sigma_p}{p_0^2}
\label{AB}
\ee
($v_0=p_0/m$ is the mean velocity of the wave packet). 
If parameter $A$ is not very big (for example, $A\sim 10$ 
for deutrons with the kinetic energy of the order of $10\,$KeV),
then the plane wave transmission formula can be used under the simple and obvious condition $B\ll 1$.
But the situation is different for {\em slow packets\/}, when parameter $A$ can be very large.
For example, for deutrons with the kinetic energy of the order of few eV or smaller 
(i.e., for temperatures of the order of $10^3\,$K), parameter $A$ can assume the values of the order of several
hundred or thousand). 

To see what can happen if condition (\ref{cond}) is broken,
we consider the family of packets 
\be
|\vf(p)|^2= \frac{{\cal N}}{\sqrt{\sigma_p}}
\exp\left[-\beta\left(\frac{\left|p-p_0\right|^2}{\sigma_p}\right)^{\gamma/2}\right],
\label{packgam}
\ee
\[
{\cal N}=\frac{\gamma[\Gamma(3/\gamma)]^{1/2}}{2[\Gamma(1/\gamma)]^{3/2}}, \qquad
\beta= \left[\frac{\Gamma(3/\gamma)}{\Gamma(1/\gamma)}\right]^{\gamma/2},
\]
where $\gamma$ is some positive constant and $\Gamma(z)$ is the Gamma function. 
In particular, $\beta=\sqrt{2}$ for $\gamma=1$ and $\beta=1/2$ for $\gamma=2$.

\begin{figure}[htb]
\includegraphics[width=0.5\textwidth]{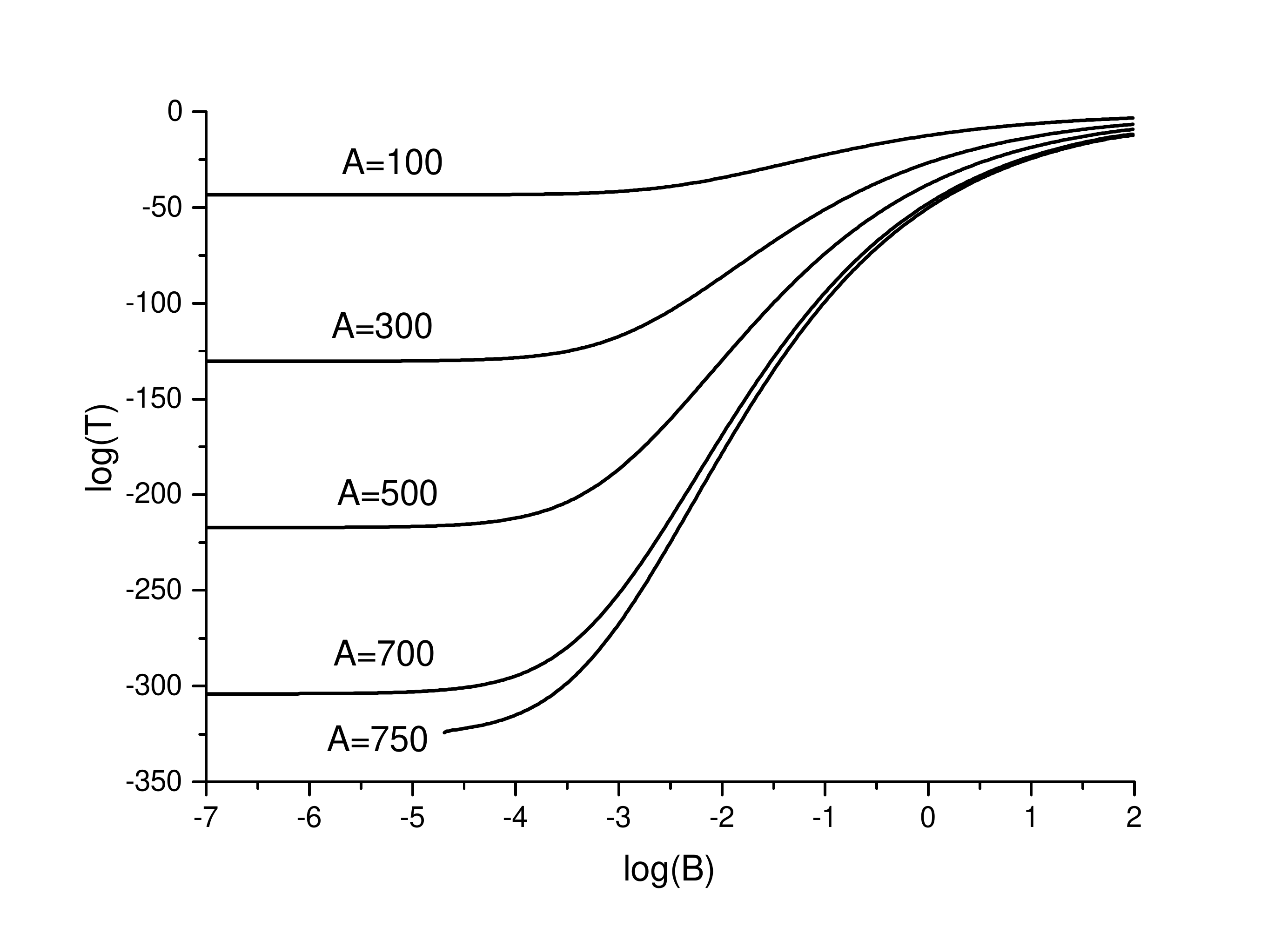}
\caption{The results of numerical calculations of the transmission coefficient (\ref{TAB}) 
for the Gaussian wave packets ($\gamma=2$).
 }
\label{fig-TAB-Gauss}
\end{figure}

\begin{figure}[htb]
\includegraphics[width=0.5\textwidth]{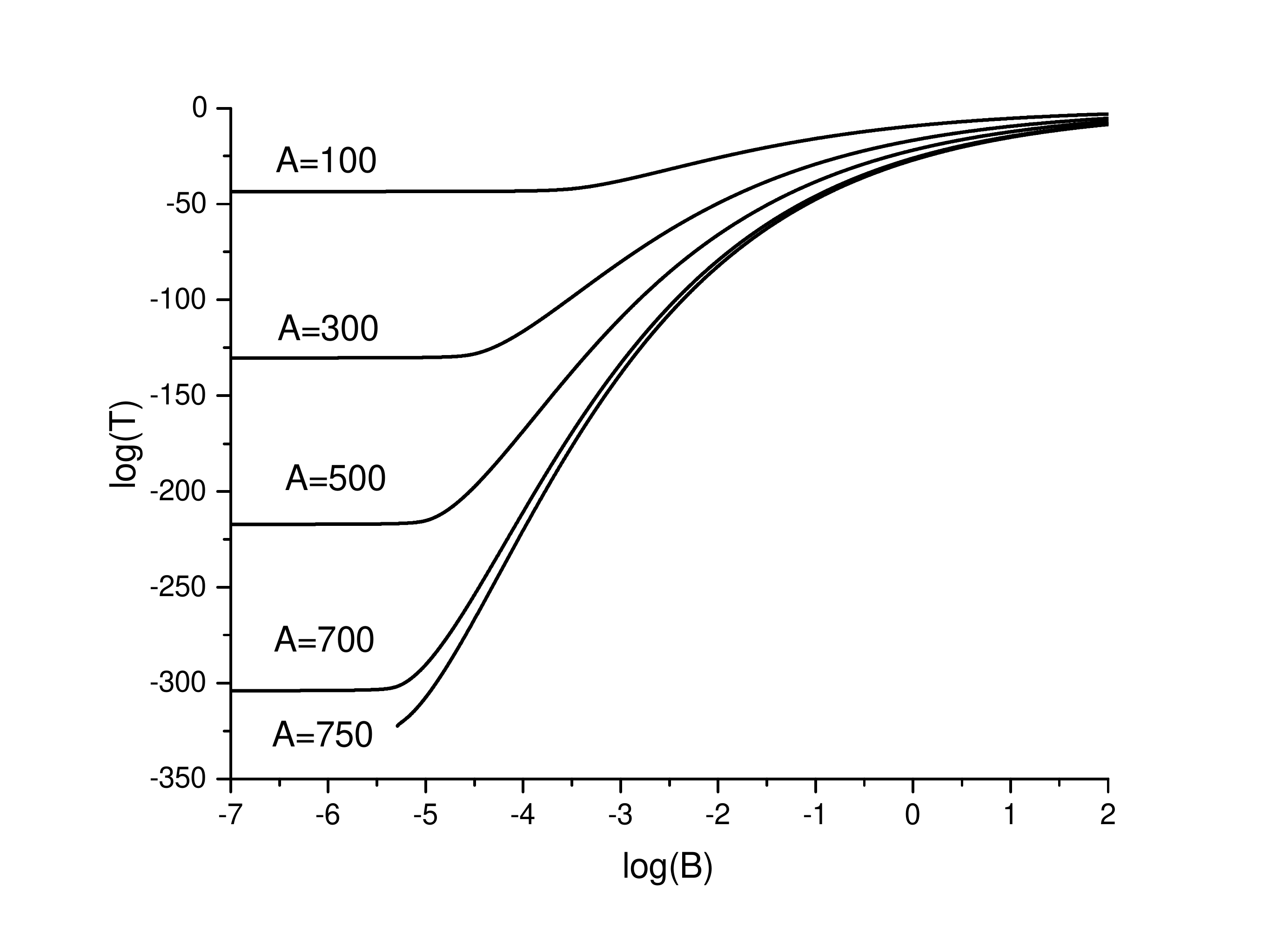}
\caption{The results of numerical calculations of the transmission coefficient (\ref{TAB})  
for the packets with $\gamma=1$.
 }
\label{fig-TAB-gamma1}
\end{figure}

Putting function (\ref{packgam}) in Eq. (\ref{Dav}), we obtain the total transmission probability
\be
T(A,B)= \frac{{\cal N}}{\sqrt{ B}} \int_0^{\infty}\exp\left[-\frac{A}{y} -
\beta\left(\frac{|y-1|^2}{B}\right)^{\gamma/2} \right] dy.
\label{TAB}
\ee
The results of numerical calculations of the integral (\ref{TAB})  for
$\gamma=2$ (Gaussian packets) and $\gamma=1$ are demonstrated in Figs. \ref{fig-TAB-Gauss}
and \ref{fig-TAB-gamma1}.
We see that  the transmission coefficient can be many orders
of magnitude bigger than the plane wave approximation value for $B\sim 10^{-3}$ (if $\gamma=2$)
and even for $B\sim 10^{-5}$ (if $\gamma=1$).

\begin{figure}[thb]
\includegraphics[width=0.5\textwidth]{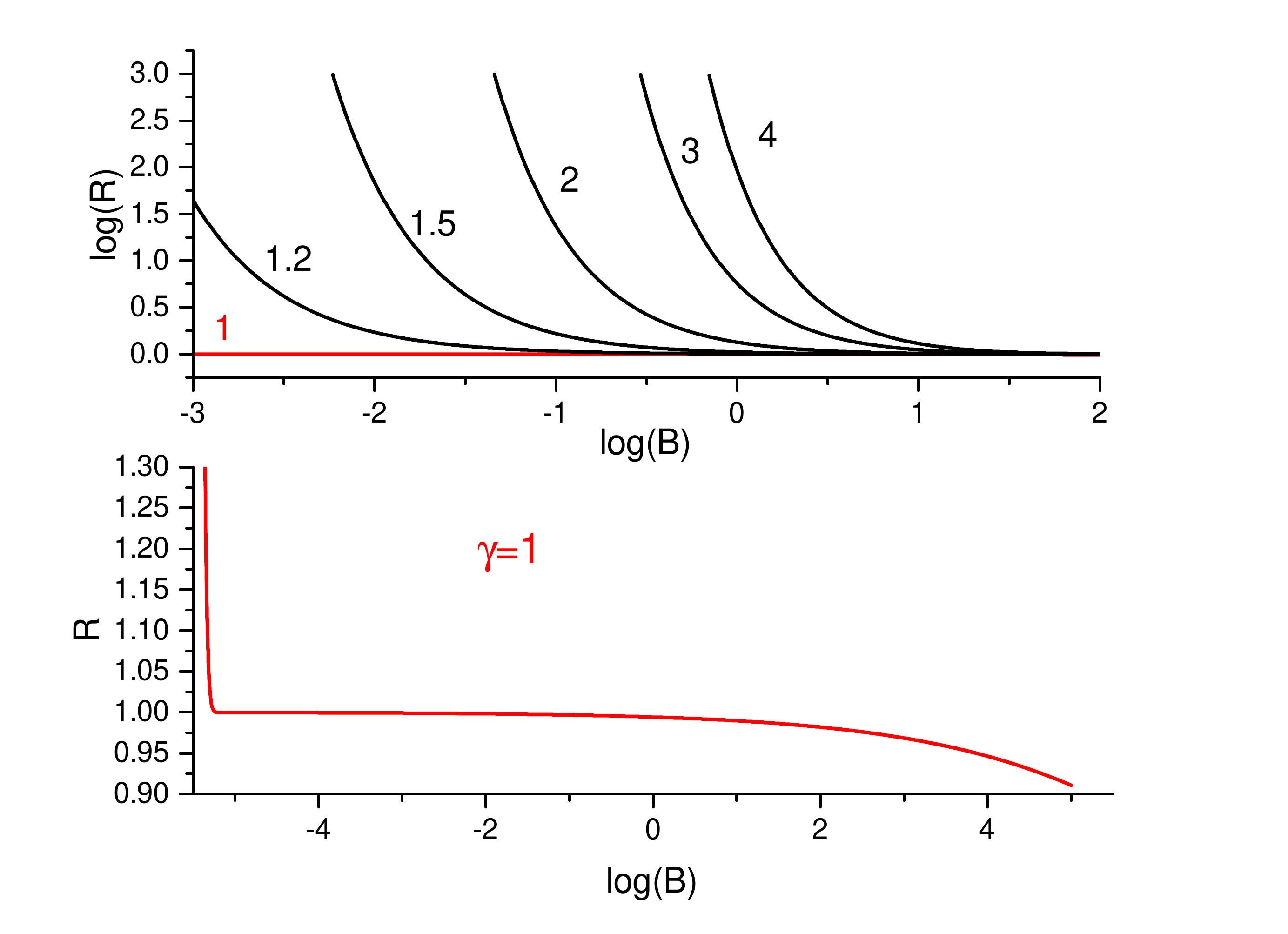}
\caption{The ratio  $R=T_*/T$ of approximate and exact numerical values of the 
transmittivity of packets with $A=700$ and different values of parameter $\gamma$.
 }
\label{fig-T*}
\end{figure}

For $A\gg 1$, we can look for an approximate analytical expression
for the integral (\ref{TAB}), using the steepest descent method. The saddle point $y_*>1$ is the solution
to the equation
\be
\frac{G}{y^2} =  (y-1)^{\gamma-1}, \qquad
G=AB^{\gamma/2}/(\gamma\beta).
\label{eq-Gy}
\ee
An approximate solution to Eq. (\ref{eq-Gy}) can be obtained for $G\gg 1$:
\be
y_* \approx G^{1/(\gamma+1)} +\frac{\gamma-1}{\gamma+1}.
\label{soly*}
\ee
Then standard formulas of the steepest descent method lead to the approximate expression
\beqn
{T}_* &\approx & {\cal N} \sqrt{\frac{2\pi}{\gamma+1}}
(\gamma\beta)^{-\frac{3}{2(\gamma+1)}}
 \left(\frac{B}{A^2}\right)^{\frac{\gamma-2}{4(\gamma+1)}}
 \nonumber \\ && \times
\exp\left[ -(\gamma\beta)^{\frac{1}{\gamma+1}} \left(\frac{A^2}{B}\right)^{\frac{\gamma}{2(\gamma+1)}}
\left(\frac{\gamma+1}{\gamma}- G^{-\frac{1}{\gamma+1}} \right)\right].
\label{T*}
\eeqn
The accuracy of formula (\ref{T*}) is illustrated in Fig. \ref{fig-T*}, where we plot the ratio
of $T_*$ to the exact numerical value $T$ as function of $\log(B)$ for the fixed value of $A=700$
but different values of parameter $\gamma$.
We see that for $\gamma>3/2$ and $A=700$, the steepest descent method gives a good approximation,
provided $B\gtrsim 1$. This is the consequence of the condition $ G^{1/(\gamma+1)}  \gg 1$:
see Eq. (\ref{soly*}). For example, for $A=700$ and $B=0.1$ we have $G=70^{1/3}\approx 4$, which is not
big number.

On the other hand, formula (\ref{T*}) gives an excellent approximation to the numerical
results for $\gamma=1$ [when  (\ref{soly*}) is the {\em exact\/} solution to Eq. (\ref{eq-Gy})],
even for very small values of parameter $B$. To understand why this happens, we notice that for
$\gamma=1$  and $A\gg 1$, the absolute value $|y-1|$ in the argument of the integral (\ref{TAB})
can be replaced by the difference $y-1$ in the whole interval $0<y<\infty$. The error due to  
incorrect values of this function in the interval $0<y<1$ is strongly suppressed by the factor
$\exp(-A/y)$ in this interval. Then integral (\ref{TAB}) can be calculated exactly, in view
of formula 3.324.1 from \cite{Grad}:
\be
\int_0^{\infty}\exp\left(-{\xi}/{y} - \eta y  \right) dy
=2\sqrt{\xi/\eta}K_1\left(2\sqrt{\xi\eta}\right),
\label{K1}
\ee
where $K_1(z)$ is the Bessel function of the third kind (the Macdonald function).
In our case, the product $\xi\eta$ is proportional to the ratio $A/\sqrt{B}$, i.e., it is
very large. Using the known asymptotical formula 
$K_1(z)\approx \sqrt{\pi/(2z)}\exp(-z)$ for $z \gg 1$ and calculating all coefficients,
we arrive at the formula
\be
T_1(A,B) \approx {\cal N}\sqrt{\pi}\left(\frac{A^2}{8B} \right)^{1/8}
\exp\left[-2\left(\frac{2A^2}{B} \right)^{1/4} +\sqrt{\frac{2}{B}} \right],
\ee
which coincides exactly with (\ref{T*}) for $\gamma=1$.

It was supposed some time ago \cite{DKM79}, that the transmission probability for the so called
correlated wave packets (with nonzero correlation coefficient between the coordinate and momentum
$r=\sigma_{xp}/\sqrt{\sigma_x\sigma_p}$) can be higher than for uncorrelated packets, and that the
increase of this probability can be described (at least approximately) by means of replacing
the true Planck constant by the effective constant $\hbar_{ef}=\hbar/\sqrt{1-r^2}$.
This question was studied later in \cite{DKM-PLA96,Vysot10,Vysot12a,Vysot13,DD-JRLR14}.  
The results of that papers show that in majority of cases, 
the nonzero correlation coefficient does increase the
probability of tunneling (although there exist specific configurations, 
where the probability can diminish \cite{DKM-PLA96}). 
But the concept of effective Planck constant seems to not work for the Coulomb potential 
(at least in the case of tunneling of free wave packets, considered in this article).
Indeed, for the Gaussian packets ($\gamma=2$)
with a fixed coordinate variance $\sigma_x$, the momentum variance increases with the correlation
 coefficient as $\sigma_p=\sigma_{p0}/\left(1-r^2\right)$.
Strongly correlated states ($r\to 1$) have large momentum variances.
In this case, Eq. (\ref{T*})  yields 
\[
T_* \sim \exp\left[-(3/2)\left(a^2/\sigma_p\right)^{1/3} \right]= 
\left[T\left(r=0\right)\right]^{\left(1-r^2\right)^{1/3}},
\]
and the last expression is rather different
from the formula 
$T(r) \sim \left[T\left(r=0\right)\right]^{\left(1-r^2\right)^{1/2}}$, suggested in
\cite{Vysot12a,Vysot13} on the basis of the concept of effective Planck constant.

\section{Conclusions}
Our main conclusion is that for high Coulomb barriers and slow particles,
it is impossible to evaluate the transmission probability without precise knowledge
 of the concrete shape of wave packet (for example, the value of parameter $\gamma$
 in the simplest cases).
The plane wave approximation can result in underestimating the real transparency by many
orders of magnitude. For some families of packets (such as (\ref{packgam}) with $\gamma \approx 1$),
the transmission coefficient can be much bigger than that given by the plane wave approximation,
even for very small ratios $B= {\sigma_p}/{p_0^2} \sim 10^{-5}$.

Note that in the regimes of validity of approximation (\ref{T*}) (especially for $\gamma \approx 1$), 
the transmission probability
depends mainly on the ratio $A^2/B= a^2/\sigma_p$, so that it depends not on the 
mean value of momentum $p_0$ or total particle energy 
$\langle E\rangle =\left(p_0^2 + \sigma_p\right)/(2m)$, but
on the momentum dispersion only. This result seems to be quite unexpected.

\section*{Acknowledgments}
The authors acknowledge a partial support of the Brazilian agency CNPq. 
We thank Prof. V. I. Man'ko for discussions that stimulated this research.

\end{document}